\documentstyle[12pt]{article} 
\newcommand{\be}{\begin{equation}}
\newcommand{\ee}{\end{equation}}
\newcommand{\bea}{\begin{eqnarray}}
\newcommand{\eea}{\end{eqnarray}}

\begin{document}

\begin{center}
{\Large\bf Parity violation in four and higher dimensional spacetime with torsion}\\[20mm]
Biswarup Mukhopadhyaya\footnote{E-mail: biswarup@mri.ernet.in}\\
{\em Harish-Chandra Research Institute,
Chhatnag Road, Jhusi, Allahabad - 211 019, India}
 
Somasri Sen \footnote{E-mail: somasri@cosmo.fis.fc.ul.pt}\\
{\em CAAUL, Departamento de Fisica da FCUL, Campo Grande, 1749-016 Lisboa, Portugal}

Soumitra SenGupta \footnote{E-mail: tpssg@iacs.res.in} \\
{\em Department of Theoretical Physics, Indian Association for the
Cultivation of Science,\\
Calcutta - 700 032, India} 

Saurabh Sur \footnote{E-mail: saurabh@juphys.ernet.in}\\
{\em Department of Physics, Jadavpur University,
Calcutta - 700 032, India}\\[20mm]
\end{center}

 
\bigskip
\abstract{ The possibility of parity violation in a gravitational theory with 
torsion is extensively explored  in four and higher dimensions. In the former  case,
we have listed our conclusions on when and whether parity ceases to be conserved, 
with both two-and three-index antisymmetry of the torsion field. In the 
latter, the bulk spacetime is assumed to have torsion, and the survival of
parity-violating terms in the  four dimensional effective action is studied, using 
the compactification schemes proposed by Arkani-Hamed-Dimopoulos-Dvali and 
Randall-Sundrum. An interesting conclusion is that the torsion-axion duality arising 
in a stringy scenario via the second rank antisymmetric Kalb-Ramond field leads to  
conservation of parity in the gravity sector in any dimension.  However, parity-violating 
interactions do appear  for spin 1/2 fermions in such theories, which can have crucial 
phenomenological implications.}

\thispagestyle{empty}

\newpage

\setcounter{page}{1}

\section{Introduction}

Torsion in space-time is an interesting possibility in theories of gravitation.
In particular, the presence of
matter fields with spin has often been suggested as a likely source of torsion. 
Ever since the Einstein-Cartan (EC) theory was proposed, the customary way to 
incorporate torsion  has been to include it as a tensorial extension to the
affine connection, which is antisymmetric in {\it at least} two indices. 
It has been further pointed out in some recent studies \cite{bmssg, bmssgss} that once 
torsion is present, a similar pseudo-tensorial extension, involving torsion and the 
completely antisymmetric tensor density, is also possible. This can, in general, cause the
violation of parity both in the pure gravity sector (including torsion) and 
in the coupling of various matter fields with torsion. 

In addition, torsion has sometimes been linked with  string theories, where 
it is possible to relate torsion to the rank-2 antisymmetric Kalb-Ramond (KR) 
field. In such a case, the field strength tensor corresponding to the KR field enters in the
connection, and it is antisymmetric in {\it all three} indices. The constraints imposed
by such complete antisymmetry requires a reappraisal of the scenario, especially
with regard to parity-violation.

The motivation of looking for parity violating gravitational interaction emerges from
both theoretical and observational viewpoints. Einstein's general relativity is known to conserve parity.
The possibility of a parity violating extension was pointed out in the usual Einstein-Cartan framework by extending the
Lagrangian density $R$, i.e the scalar curvature, by $R + \epsilon^{\mu\nu\alpha\beta} R_{\mu\nu\alpha\beta}$ ,which is the 
only possible extension linear in $R$. Although this new term vanishes identically for Einstein's theory but it yields a 
non-vanishing parity violating contribution for Einstein-Cartan theory. In ref.\cite{bmssg,bmssgss} it has been pointed out that such a 
parity violating term comes naturally if one considers a pseudo tensorial extension of the affine connection.In fact there is
no apriori reason to beleive that the Cartan extension of the affine connection must have a definite parity i.e. 
parity conserving only.
Thus getting parity violation in this way looks more natural. In addition this allows us to study the coupling of
this parity violating term ( pseudo tensorial extension of the affine connection ) with other spin fields through
the usual minimal coupling prescription \cite{bmssg, bmssgss}.
The observational motivation emerges from two important results reported in \cite{ssgas, kamlue}.
In ref.\cite{ssgas} it has been shown that a parity violating gravitational interaction can flip
the helicity of a fermion and thereby provides a possible explanation of the well known 
neutrino anomaly problem. On the other hand ref.\cite{kamlue} shows that a parity violating coupling 
between electromagnetic and a scalar field can explain the recently observed anisotropy in the 
Cosmic Microwave Background ( CMB ) radiation. Indeed the dual scalar of the pseudo-tensor component of the connection 
discussed above can be identified with such a scalar.

Further investigations have been recently carried out in the context of theories with
large extra dimensions, namely those of Arkani-Hamed--Dimopoulos--Dvali (ADD) \cite{add} 
and Randall--Sundrum (RS) \cite{rs}. In such models 
torsion exists in the bulk together with gravity, while all the standard model fields
are confined to a 3-brane. It has been demonstrated \cite{bmssgsen}, that a bulk torsion 
related to the KR field in ADD type of models has most of its parity-violating effects 
washed out when one compactifies the extra dimensions and considers the projection 
of bulk fields on the visible brane. In the context of a Randall-Sundrum scenario, very 
similar conclusions hold in the simplest cases. However, there one reaches the interesting 
conclusion that in spite of having the same status in the bulk as gravity, effects of the
massless mode arising from torsion are heavily suppressed on the standard model 
brane, thus creating the illusion of a torsionless universe \cite{bmsenssg}.

On the whole, the issue of parity-violation in torsioned gravity needs to be addressed 
in the light of a number of factors, namely

\begin{itemize}
\item Whether the extension due to the torsion field has antisymmetry in
{\it two} or {\it three} indices.

\item Whether the coupling constants corresponding to the different pseudo-tensorial 
extensions are the same or different.

\item The dimensionality of the space in which torsion is assumed to exist. 

\item Whether torsion is introduced minimally (i.e. through the terms of least order) or 
non-linear extensions are to be made if one considers the possibility of parity violation. 
\end{itemize}  

In this paper, we present our observations for different cases arising out of 
combinations of the above possibilities. Although some of the individual points have 
been discussed earlier in the references given above, an overall perspective is yet 
to be provided on this unique feature of gravitational interactions. Such a perspective
is aimed at in this work. 

In section 2, we outline the general features of the mechanism of parity-violation induced
by torsion. An examination of individual cases in both 4 and higher dimensions, with the 
ways the parity-violating terms can be constructed in each case, is made in section 3.
We summarise and conclude in section 4.

\section{Torsion and parity violation} 

\subsection{The framework}

The connection in EC theory, denoted by $\tilde{\Gamma}^{\mu}_{\nu \lambda}$,
 is completely asymmetric in all its indices. Upon antisymmetrization 
of $\tilde{\Gamma}^{\mu}_{\nu \lambda}$ in 
the two lower indices $\nu$ and $\lambda$, one obtains a tensor known as `spacetime torsion':
\be
H^{\mu}_{~\nu \lambda} ~=~ \frac 1 2 \left(\tilde{\Gamma}^{\mu}_{\nu \lambda} 
~-~ \tilde{\Gamma}^{\mu}_{\lambda \nu}\right).
\ee
Accordingly, $\tilde{\Gamma}^{\mu}_{\nu \lambda}$ can be expressed in terms of
the symmetric Christoffel connection $\Gamma^{\mu}_{\nu \lambda}$ and the torsion as
\be
\tilde{\Gamma}^{\mu}_{\nu \lambda} ~=~ \Gamma^{\mu}_{\nu \lambda} ~-~
K^{\mu}_{~\nu \lambda}
\ee
where $K^{\mu}_{~\nu \lambda} = H^{\mu}_{~\nu \lambda} + H^{~\mu}_{\lambda~\nu}
- H^{~~\mu}_{\nu \lambda}$ is known as the  `contorsion' tensor, constructed out of 
the torsion tensor in order to preserve the metricity condition in EC theory:
\be
\tilde{D}_{\nu}~ g^{\mu \nu} ~=~ 0,
\ee
$\tilde{D}$ being the covariant derivative defined in terms of $\tilde{\Gamma}$.
The contorsion tensor is, by construction, antisymmetric in the first and the third
covariant (contravariant) indices.

A straightforward way to introduce parity-violation through the well-known minimal 
coupling scheme is to incorporate a pseudo-tensorial extension in the EC connection 
\cite{bmssg} such that
\be
\tilde{\Gamma}^{\mu}_{\nu \lambda} ~\rightarrow~ \tilde{\Gamma}^{\mu}_{\nu \lambda}
~=~ \Gamma^{\mu}_{\nu \lambda} ~-~ \left(H^{\mu}_{~\nu \lambda} ~+~ H^{~\mu}_{\lambda
~\nu} ~-~ H^{~~\mu}_{\nu \lambda} \right) ~-~ q~\left(^{*}H^{\mu}_{~\nu \lambda}
~+~ ^{*}H^{~\mu}_{\lambda~\nu} ~-~ ^{*}H^{~~\mu}_{\nu \lambda}\right)
\ee
with ~$^{*}H^{\mu}_{~\nu \lambda}$~ having opposite parity properties relative to
~$H^{\mu}_{~\nu \lambda}$. The parameter $q$ determines the degree
of parity-violation, and as a general notation we are using the ~$*$~ for a
pseudo-tensor. In general, ~$H$~ and ~$^{*}H$~ may be two completely different 
tensors transforming oppositely under spatial parity, but in that case it is always 
possible restore parity through appropriate rephasing of the fields. 
Therefore, the only situation where one can have a parity-violating gravitational 
field theory is when ~$^{*}H$~ is constructed out of ~$H$~ itself and linearly in
the completely antisymmetric permutation tensor ~$\epsilon$.~ For example, in 4 
dimensions, a valid combination of ~$\epsilon$~ and ~$H$~ creating a ~$^{*}H^{\mu}_
{\nu \lambda}$~ (antisymmetric in $\nu, \lambda$) may be ~$\epsilon^{\alpha \beta}_
{~~\nu \lambda} H^{\mu}_{~\alpha \beta}$~ or, ~$\epsilon^{\mu \alpha}_{~~\beta 
[\nu} H^{\beta}_{~\lambda] \alpha}$,~ as is shown in \cite{bmssg,bmssgss}. 

\subsection{$H$ with two-index antisymmetry}
As has been mentioned above, ~$H$~ is antisymmetric in two indices in the most general case.
If parity has to be violated, then a similar general property has to be attributed to
~$^{*}H$~ as well, since the latter is constructed linearly out of the former in
a minimal construction. In such a case, the gravitational 
Lagrangian density, with the surface terms eliminated, turns out to be composed 
of two parts transforming oppositely under parity. The parity conserving part 
${\cal L}_{grav}^{~~(pc)}$ and the parity violating part ${\cal L}_{grav}^{~~(pv)}$ are
given as
\bea
{\cal L}_{grav}^{~~(pc)} ~=~ R(g) ~-~ H^{\mu}_{~\nu \lambda} \left(
H_{\mu}^{~\nu \lambda} ~-~ 2 H_{~~\mu}^{\nu \lambda} \right)
~-~ H^{\alpha}_{~\alpha \beta} H_{\mu}^{~\mu \beta}~+~ O(q^2) \\
{\cal L}_{grav}^{~~(pv)} ~=~ - 2 q \left( H^{\mu}_{~\nu \lambda} ~
^{*}H_{\mu}^{~\nu \lambda} ~-~ H^{\mu}_{~\nu \lambda}~ ^{*}H_{~~\mu}^{\nu 
\lambda} ~-~ ^{*}H^{\mu}_{~\nu \lambda}~ H_{~~\mu}^{\nu \lambda} 
~+~ 2 H^{\alpha}_{~\alpha \beta}~ ^{*}H_{\mu}^{~\mu \beta} \right).
\eea
where $O(q^2)$ are the additional parity-conserving terms arising
in the present scenario; they are of less relevance since we are
primarily interested in the terms bearing opposite parity properties
relative to the original Cartan terms. 
 
It should also be mentioned here that ${\cal L}_{grav}^{~~(pv)}$ above is
identical to the form proposed in an earlier work \cite{hoj} where
an extra term of the form ~$\epsilon^{\alpha \beta \mu \nu}~R_{\alpha \beta \mu 
\nu}$~ was added to the original Einstein-Hilbert Lagrangian.
However, the present scheme gives us in addition a guideline for constructing
parity violating terms involving matter fields with different spins.

For a spin-1/2 fermion in a spacetime with torsion, the extended Dirac Lagrangian 
density is given by \cite{aud}:
\be
{\cal L}_{tor}^{f} ~=~ \bar{\psi}~\left[i \gamma^{\mu}~\left( \partial_{\mu} ~-~
\sigma^{\rho \beta} v^{\nu}_{\rho} g_{\lambda \nu} \partial_{\mu} v^{\lambda}_{\beta}
~-~ g_{\alpha \delta} \sigma^{a b} v^{\beta}_a v^{\delta}_b \tilde{\Gamma}^{\alpha}
_{\mu \beta} \right) \right]~\psi
\ee
where ~$v^{\mu}_a$~ denotes the tetrad connecting the curved space with the
corresponding tangent space. The above expression can be decomposed into the terms
with opposite parity:
\bea
{\cal L}_{tor}^{f~(pc)} ~=~ {\cal L}_{E}^{f} ~-~ \bar{\psi}~\left[i \gamma^c~g_{\alpha \delta} 
\sigma^{a b} v^{\mu}_c v^{\beta}_a v^{\delta}_b~ \left(H^{\alpha}_{~\mu \beta} ~+~ 
H^{~\alpha}_{\beta~\mu} ~-~ H^{~~\alpha}_{\mu \beta}\right)  \right]~\psi \\
{\cal L}_{tor}^{f~(pv)} ~=~ - q ~\bar{\psi}~\left[i \gamma^c~g_{\alpha \delta} \sigma^{a b} 
v^{\mu}_c v^{\beta}_a v^{\delta}_b~ \left(^{*}H^{\alpha}_{~\mu \beta} ~+~ 
^{*}H^{~\alpha}_{\beta~\mu} ~-~ ^{*}H^{~~\alpha}_{\mu \beta}\right)  \right]~\psi
\eea
${\cal L}_{E}^{f}$~ being the Dirac Lagrangian density in Einstein gravity.
Thus explicit parity-violation appears through the term ${\cal L}_{tor}^{f~(pv)}$ when a 
spin - 1/2 fermion couples to the background torsion. Just a two-index antisymmetry
in the torsion tensor is thus sufficient to ensure parity violation in both the
pure gravity sector and in the Lagrangian of spin-1/2 particles.

The coupling of torsion with a spin-1 Abelian gauge field $A_{\mu}$, however, 
runs into problems in maintaining gauge invariance.
This is because the corresponding field strength
$\tilde{F}_{\mu \nu} = \tilde{D}_{[\mu} A_{\nu]}$ is not invariant under $U(1)$ gauge
transformation. This has been a persistent difficulty for torsion
with two-index antisymmetry,  so long as one wants to remain within the
minimal coupling scheme. In a string theoretic scenario, however,
this problem can be handled in a manner to be discussed below.

\subsection{$H$ with three-index antisymmetry}

A torsion tensor $H$ with complete antisymmetry in all its indices 
may be identified with the field strength corresponding to 
the second rank antisymmetric tensor field $B_{\mu\nu}$ 
appearing in the massless sector of heterotic string theory. 
Starting from the Einstein-Cartan action in such an antisymmetric tensor field 
background one can use the equation of motion for torsion to identify torsion with the KR field strength
and trade away the torsion from the action \cite{pmssg}. 

To cancel $U(1)$ 
gauge anomaly and preserve N=1 supersymmetry in the heterotic string theory the field strength $H_{\mu \nu \lambda}$ is 
augmented suitably with
a Chern-Simons (CS) term $A_{[\mu} F_{\nu \lambda]}$ ($F$ being the field strength
of a $U(1)$ gauge field $A$): 
\be
H_{\mu \nu \lambda} ~=~ \partial_{[\mu} B_{\nu \lambda]} ~+~ A_{[\mu} \partial_{\nu}
A_ {\lambda]}
\ee  
Using this Chern-Simons augmented field strength $H_{\mu \nu \lambda}$ we consider the 
low energy field theory limit of the bosonic sector of
the toroidally compactified String theory. 
It has been shown \cite{pmssg} that in such a theory CS term plays the crucial role in 
resolving the problem of $U(1)$ gauge-invariance mentioned above. This 
is easy to verify since $H_{\mu \nu \lambda}$ as defined above is invariant under 
the $U(1)$ gauge-transformation $\delta A_{\mu} = \partial_{\mu} \omega$ and  
$\delta B_{\mu \nu} ~=~ \omega~ F_{\mu \nu}$ \cite{pmssg}.

Now, one can again have a pseudo-tensor ~$^{*}H$~ constructed out of the permutation 
tensor $\epsilon$ and $H$ and write in general the torsion as ~$H_{\mu \nu \lambda} 
~+~ q~ ^{*}H_{\mu \nu \lambda}$. The sum as a whole need not be totally antisymmetric, 
as ~$^{*}H$~ can be antisymmetric only in a pair of indices although 
$H_{\mu \nu \lambda}$ is antisymmetric in 
all the three indices. Such construction is explicitly
shown in the following section, where we shall also state the specific
conditions for retaining parity violating effects in different sectors.
Due to the presence of the CS term,  the 
Einstein-Cartan-Kalb-Ramond (ECKR)  Lagrangian density  
involves the gauge field $A$. Therefore, following the formalism in \cite{pmssg}
we can express the Lagrangian density for ECKR-gauge field coupling as
\be
{\cal L}_{ECKR}^{gauge} ~=~ R(g) ~-~ \frac 1 {12} (H^{\mu \nu \lambda}
~+~ q~^{*}H^{\mu \nu \lambda}) (H_{\mu \nu \lambda} ~+~ q~^{*}H_{\mu \nu \lambda})
~-~ \frac 1 4 F_{\mu \nu} F^{\mu \nu}  
\ee

The Lagrangian density for ECKR-fermion coupling is given by
\be
{\cal L}_{ECKR}^f ~=~ {\cal L}_{E}^{f} ~-~ \bar{\psi} \left[i \gamma^c g_{\alpha \delta} 
\sigma^{a b} v^{\mu}_c v^{\beta}_a v^{\delta}_b \left\{H^{\alpha}_{~\mu \beta} ~+~
q \left(^{*}H^{\alpha}_{~\mu \beta} ~+~ ^{*}H^{~\alpha}_{\beta~\mu} ~-~ 
^{*}H^{~~\alpha}_{\mu \beta} \right) \right\} \right] \psi
\ee

\section{Construction of ~$^{*}H$~ in different dimensions}

\subsection{The general outlook}

So far we have discussed in a general way the possibility of parity 
violation arising from $^{*}H$. Now we shall 
concentrate on various ways of constructing $^{*}H$ out of $H$ in different 
spacetime dimensions.  

Depending on the dimensionality, ~$^{*}H$~ can be constructed using 
linear as well as higher powers of ~$H$.~ In particular, it is straightforward 
to see that 

(a) In even spacetime dimensions (4, 6 $\cdots$) ~$^{*}H$~ must be constructed 
using an odd number of ~$H$'s,~i.e., $^{*}H$ is linear, cubic $\cdots$ in $H$.

(b) In odd spacetime dimensions (5, 7, $\cdots$) ~$^{*}H$~ must contain an even 
number of ~$H$'s~ and therefore can be bilinear, quadrilinear $\cdots$ in $H$. 
 
Note that since the three-form ~$H$~ is equal to ~$d B + A \wedge F$, dimensional 
arguments tell us that an ~$^{*}H$~ constructed out of higher powers of ~$H$'s
gives parity-violating effects suppressed by correspondingly higher powers of 
the Planck mass. 

\subsection{Construction of ~$^{*}H$~ in 4 dimensions}
\subsubsection{Minimal construction}

We are now considering an $H$ which is totally antisymmetric in all three indices.
In 4 dimensions, a minimally constructed $^{*}H$~ consists of terms linear in ~$H$,~ 
whence the pseudo-tensorial connection can generically be written as \cite{bmssgss} 
\be
q~^{*}H^{\mu}_{~\nu \lambda} ~=~ q_{1}~\epsilon^{\alpha \beta}_{~~\nu \lambda}~ 
H^{\mu}_{~\alpha \beta} ~+~ q_{2}~\epsilon^{\mu \sigma}_{~~\rho [\nu}~
H^{\rho}_{~\lambda] \sigma}.
\ee
However, if the coupling strengths $q_1$ and $q_2$ are equal (which is the situation 
corresponding to complete antisymmetry in the pseudo-connection), the above expression 
vanishes identically as a whole.
This can be verified easily on observing that one can always replace the 
totally antisymmetric three-tensor ~$H_{\mu \nu \lambda}$~ --- a three-form 
--- by its  {\it Hodge}-dual one-form, i.e., a pseudo-vector ~$h_{\sigma}$ as:
\be
H_{\mu \nu \lambda} ~=~ \epsilon_{\mu \nu \lambda \sigma}~h^{\sigma}. 
\label{dual}
\ee
Therefore $q_1$ and $q_2$ must always differ, which implies that we are left 
with the case where the term ~$^{*}H^{\mu}_{~\nu \lambda}$~ is 
antisymmetric in $\nu$ and $\lambda$ only. This is not an unnatural
assumption, since there is no symmetry of the theory demanding the
equality of the two charges.

Even with $q_1$ and $q_2$ unequal, a rather interesting thing is observed.
If one considers the parity violating part of the gravity sector (equation 6)
and uses the above duality relation, it is straightforward to see that 
${\cal{L}}_{grav}^{(pv)}~=~0$ identically. Thus a Kalb-Ramond type of torsion 
cannot violate parity in the effective scalar curvature.

A similar conclusion follows for Abelian gauge fields, too.
The gauge-invariant ECKR-Lagrangian density along with the gauge field $A$
[Eq.(11)] can now be separated into parity conserving (pc) and parity violating (pv) parts as
\bea
{\cal L}_{ECKR}^{gauge~(pc)} &=& R(g) - \frac 1 4 F_{\mu \nu} F^{\mu \nu}  
- \frac 1 {12} (\partial_{[\mu} B_{\nu \lambda]} + A_{[\mu} F_{\nu\lambda]}) 
(\partial^{[\mu} B^{\nu \lambda]} + A^{[\mu} F^{\nu\lambda]}) + O(q_1, q_2)^2\\
{\cal L}_{ECKR}^{gauge~(pv)} &=&-~\frac 1 6 ~(q_1 ~+~ 2 q_2) ~\epsilon^{\rho \alpha}_
{~~\beta \sigma}~(\partial_{[\lambda} B_{\rho \alpha]} + A_{[\lambda} F_{\rho
\alpha]}) (\partial^{[\lambda} B^{\beta \sigma]} + A^{[\lambda} F^{\beta\sigma]})
\eea

However, once again the relation (14) can be employed to check that the  
parity-violating term ~${\cal L}_{ECKR}^{gauge~(pv)}$~ vanishes identically.This is
because all the terms, including those from Chern-Simons extension are three index antisymmetric
and therefore dual to a vector in four dimension by the relation (14). 
So our conclusion is that the theory is parity conserving in both gravity 
and electromagnetic sectors even for $q_1 \neq q_2$.

It is worth mentioning here that in a recent work \cite{pm}, it has
been argued that there can be an alternative way of incorporating 
parity-violation in the coupling of the gauge field with torsion by extending the 
Chern-Simons term in the modified field strength $H_{\mu\nu\lambda}$ by the 
dual of Maxwell field tensor $F^{\mu\nu}$. Such a term generates a parity 
violating interaction between the gauge field and the torsion.

Once the torsion tensor is identified with the KR field, the pseudo-tensorial
extension of the connection makes the KR coupling to a spin - 1/2 fermion 
parity-violating. In terms of the {\it axion} that appears in the string spectrum and
defined through the duality relation
\be
\partial_{[\mu} B_{\nu \lambda]} ~=~ \epsilon_{\mu \nu \lambda \sigma}
~\partial^{\sigma}~\phi.
\ee
the Lagrangian density in the fermionic sector is given by
\bea
{\cal L}_{ECKR}^{f} &=&  {\cal L}_{E}^{f}~+~8~(q_1 ~-~ q_2) ~\bar{\psi}~\left( i \gamma_c~ 
\sigma^{a b} ~v^{\lambda}_a v^{\mu}_b v^{\nu}_{c} ~g_{\nu\lambda}~\partial_{\mu}
~\phi \right)~\psi \nonumber \\
&+& \bar{\psi}~\left( i \gamma_c~ \sigma^{a b} ~v^{\lambda}_a v^{\mu}_b 
v^{\nu}_{c} \left[ 2 q_1 ~\epsilon^{\alpha \beta}_{~~\nu \lambda}~ A_{[\alpha}
F_{\beta\mu]}  ~-~ (q_1 - 2 q_2) ~\epsilon^
{\alpha \beta}_{~~\mu \nu}~A_{[\alpha} F_{\beta\lambda]} 
\right]\right)~\psi 
\eea
This Lagrangian density is manifestly parity-violating, through both 
the axion $\phi$ and the CS term. Thus
fermionic interactions constitute the benchmark of parity
violation induced by torsion of the above kind, albeit with the special 
requirement $q_1~\neq~q_2$.
Moreover, due to the presence of the CS term in the connection, an 
interaction between the gauge field and the fermion appears. Though 
the term is suppressed by two powers of the 
Planck mass, such an interaction may be interesting 
for future studies.

Before we move on to the next topics, we summarize below our main
conclusions on parity violation with torsion in four
dimensions, with the pseudo-tensorial extension always  kept linear
in $H$:

\begin{itemize}
\item When the torsion tensor is antisymmetric {\em only in the two lower indices}
({\it i.e.} in a model-independent representation of torsion), parity
violation is always observed both in the pure gravity sector ({\it i.e.} in
the effective scalar curvature) and in the coupling of matter fields to 
torsion. However, the coupling of torsion to massless gauge fields is difficult 
to ensure unless one goes beyond the minimal scenario.

\item When the torsion tensor is antisymmetric in {\em all three indices},
({\it i.e.} one can write it both as the strength of the antisymmetric
Kalb-Ramond tensor field and and as the dual of a pseudoscalar field),
the pseudotensorial extension ${^*}H$ identically vanishes so long as it is also
constructed as {\em antisymmetric in all three indices}. Thus there is no
possibility of parity violation in such a case.

\item When the torsion tensor is antisymmetric in {\em all three indices},
it is still possible to have only a {\it two-index antisymmetry} in the 
pseudo-tensorial part ${^*}H$, by imposing inequality of the 
two couplings $q_1$ and $q_2$. In such a case, too, the gravity
sector and the gauge field sector still turn out to be parity-conserving.
However,spin-1/2 fields have parity-violating coupling with torsion in such 
a case. 
\end{itemize}

\subsubsection{Non-minimal construction}

We have already seen that no parity violation occurs in the gauge and gravity 
sectors for the minimal (linear) extension in 4 dimensions. Here we look for the 
possibility of parity violation in these sectors with the leading non-minimal terms in the 
extension. As mentioned earlier, in 4 dimensions, a pseudo-tensor ~$^{*}H$~ constructed
using ~$H$~ can, in general, have terms containing only odd powers of ~$H$. 
Therefore, the most general pseudo-tensorial part of the affine connection can be
schematically expressed as
\be
^{*}H ~=~ \epsilon H ~+~ \epsilon~ H H H ~+~ \epsilon~ H H H H H ~+~ \cdots.
\ee
The set of possible terms trilinear in $R$ (suppressing the charges multiplying the various 
terms) in the above expression is given by
\bea
\epsilon^{\alpha \beta}_{~~\nu \lambda}~ H^{\mu}_{~\alpha \rho}~
H^{\sigma \kappa}_{~~\beta}~ H^{\rho}_{~\sigma \kappa} ~+~ 
\epsilon^{\alpha \beta}_{~~\rho [\nu}~ H^{\sigma}_{~\lambda] \kappa}~
H^{\rho \kappa}_{~~\sigma}~ H^{\mu}_{~\alpha \beta} ~+~ 
\epsilon^{\alpha \mu}_{~~\beta \rho}~ H^{\beta \sigma}_{~~[\nu}~
H^{\kappa}_{~\lambda] \alpha}~ H^{\rho}_{~\sigma \kappa} \nonumber\\
~+~ \epsilon^{\alpha \mu}_{~~\beta \rho}~ H^{\beta}_{~\nu \lambda}~
H^{\rho \sigma}_{~~\kappa}~ H^{\kappa}_{~\alpha \sigma} ~+~~ similar ~terms~ 
\eea

In the special case of a completely antisymmetric pseudo-tensorial connection,
there are a number of allowed terms for each non-minimal order construction.
However, similar to the minimal construction case, terms of each variety in the
non-minimal construction can be shown to vanish on using the general relation (\ref{dual}).
Thus we can make the following generic statement: 
{\it it is, in no way, possible in a 4-dimensional Lagrangian to have a pseudoscalar 
term built out of completely antisymmetric three-tensors raised to any index}.

When the field $H$ is only two index antisymmetric then the non-minimal extensions
no longer vanish. However,  it can be explicitly checked that no parity-violating 
term in the Lagrangian density  in 4 dimensions appears  either in the gravity sector 
or in coupling with gauge fields upto a term trilinear in $H$ in the connection.
In the fermionic sector, parity violating terms from the non-minimal extensions do 
appear in the Lagrangian density. However such terms are hardly of any significance 
as they are suppressed by increasingly higher powers of Planck mass.

\subsection{Construction of ~$^{*}H$~ in 5 dimensions}

Considering that torsion (or, equivalently, the KR field) coexists alongside gravity
in the bulk, we find that in 5-dimensional spacetime the pseudo-tensor ~$^{*}H$~ 
constructed from ~$H$~ have to be at least bilinear in the latter. The most general 
pseudo-tensorial part of the affine connection, antisymmetric in a pair of indices, 
can now be written as
\bea
^{*}H^{\mu'}_{~~\nu' \lambda'} ~&=&~ \left(q_1~ \epsilon_{\alpha' \beta'
\gamma' \nu' \lambda'} H^{\mu' \alpha' \delta'} ~+~ q_2~ 
\epsilon^{\mu'}_{~~\alpha' \beta' \gamma' [\nu'} H_{\lambda']}^{~~\alpha' \delta'} 
\right) H^{\beta' \gamma'}_{~~~~\delta'} \nonumber \\
&+&~ \left(q_3~ \epsilon^{\mu'}_{~~\alpha' \beta' \gamma' \delta'} H_{\nu' 
\lambda'}^{~~~~\alpha'} ~+~ q_4~ \epsilon_{\alpha' \beta' \gamma' \delta' [\nu'} 
H_{\lambda']}^{~~\mu' \alpha'} \right) H^{\beta' \gamma' \delta'} \nonumber\\
&+&~  q_5 ~\epsilon_{\alpha' \beta' \gamma' \delta' [\nu'} 
H_{\lambda']}^{~~\alpha' \beta'} H^{\mu' \gamma' \delta'}
\eea
where the primed indices
~$\mu', \nu', \cdots$, etc. run all over both the usual four-dimensional
spacetime and the extra space dimension $y$. 
The coupling strengths $q_1, q_2, q_3, q_4$ and $q_5$ are, in general, 
different from each other, thereby leaving ~$^{*}H$~ to be antisymmetric in two indices.
The special case of a totally antisymmetric pseudo-tensor
~$^{*}H$~ can be encountered if we set $q_1 = - q_2,~q_3 = q_4$ and put $q_5 = 0$. 
Unlike in 4 dimensions, here the totally antisymmetric ~$^{*}H$~ gives 
non-vanishing contribution to the connection.

With this modified connection in 5 dimensions, we now examine the parity 
violating effect in the effective 4-dimensional theory with two compactification 
mechanisms, viz., Arkani-Hamed--Dimopoulos--Dvali (ADD) \cite{add} and 
Randall--Sundrum (RS) \cite{rs}. We compute the parity violating part of the 
4-dimensional  Lagrangian density for these two compactification schemes. For 
the sake of convenience we consider here only the terms multiplying $q_1$ and 
$q_2$ terms of Eq.(21), with $q_1 \neq q_2$ in general. The conclusions are, 
however, not affected by this simplification.  

\subsubsection{Compactification in ADD scenario}

Although the ADD type of models are phenomenologically disfavoured in 5 dimensions,
we include it here for completeness. In such a model \cite{add}, the compact and 
Lorenz degrees of freedom can be factorized and the string scale $M_s$ (expected 
to lie between a few TeV's and a few tens of TeV) controls the strength of 
gravity in ($4 + n$) dimensions. $M_s$ is related to the 4-dimensional Planck scale 
$M_p$ by~ $M_s^{n+2}/M_p^2 ~\sim~ R^{-n},~R$
being the compactification radius. Compactification
of the $n$ extra dimensions leads to a tower of Kaluza-Klein (KK) modes on a visible
3-brane and as such a massless field in the bulk gives rise to a massive spectrum
$m_{\vec{n}}^2 = 4 \pi^2 \vec{n}^2/R^2$ with $\vec{n} = (n_1, n_2, \cdots, n_n)$ 
\cite{han}. In a physical process, the summation over these tower of fields, 
convoluted with the corresponding density of states, causes an enhancement, 
despite an $M_p$-suppression
of individual coupling. Thus `new physics' is found to intervene at the TeV scale, 
thereby providing a natural cut-off to the electroweak theory.

For a bulk KR field $B_{\mu' \nu'}$, the ADD compactification in general gives rise
to a set of tensor fields $B_{\mu\nu}^{\vec{n}}$, vector fields $B_{\mu}^{\vec{n}}$
and scalar fields $\chi^{\vec{n}}$ in a 4-dimensional effective theory. However, one 
can assume the bulk $B_{\mu' \nu'}$ to be block-diagonal in compact and
non-compact dimensions \cite{bmssgsen}, i.e., $B_{\mu}^{\vec{n}}$ can be taken to be
zero without any loss of generality. Now, following the standard toroidal
compactification procedure shown in \cite{han}, we obtain the 4-dimensional 
effective parity-violating part of the Lagrangian density for KR-fermion coupling 
\bea
{\cal L}^{~(pv)}_{f} &=& 2 q_1~ \bar{\psi}~ [\sum_{n, n', m, m'} i \gamma_c
\sigma_{a b} v^a_{\mu} \epsilon^{\alpha \beta b c}~ \{ \frac{2 \pi i} R~n~ g^{\mu \rho (m)}
~g^{\nu \sigma (m')}~(B_{\nu \rho}^{(n)}~ \partial_{[\alpha} B_{\beta \sigma]}^{(n')}
\nonumber \\
&+& 2 B_{\sigma \beta}^{(n)}~ \partial_{[\rho} B_{\alpha \nu]}^{(n')}) ~-~
\frac{4 \pi^2}{R^2}~n n' ~g^{\mu \rho (m)} ~\zeta^{\nu (m')}~ (B_{\nu \rho}^{(n)}~
B_{\alpha \beta}^{(n')} ~+~ 2 B_{\rho \alpha}^{(n)}~ B_{\sigma \beta}^{(n')}) \} ]~\psi
\nonumber \\
&+& \frac{q_1 + 2 q_2} 2~ \bar{\psi}~ [\sum_{n ,n', m} \gamma^c \sigma^{a b}
\gamma^5 v^{\alpha}_a v^{\beta}_b v^{\lambda}_c ~ \{ \frac{2 \pi i} R~n~ g^{\nu \sigma (m)}
~(B_{\nu \lambda}^{(n)}~ \partial_{[\alpha} B_{\beta \sigma]}^{(n')} \nonumber\\
&+& 2 B_{\sigma \beta}^{(n)}~ \partial_{[\lambda} B_{\alpha \nu]}^{(n')})
~-~ \frac{4 \pi^2}{R^2}~n n' ~\zeta^{\sigma (m)}~ (B_{\sigma \lambda}^{(n)}~
B_{\alpha \beta}^{(n')} ~+~ 2 B_{\lambda \alpha}^{(n)} ~B_{\sigma \beta}^{(n')}) \} ]~\psi
\eea
where $\zeta^{\sigma (n)} = g^{\sigma y (n)}$ (where $y$ stands for the extra dimensions),
and with the contributions due to the CS terms which are suppressed by higher powers of
Planck mass in the above expression.

\subsubsection{Compactification in RS scenario}

In the RS framework, we have a non-factorizable geometry and as such the metric 
contains a so-called `warp'-factor which is an exponential function of the compact
space dimension $y$:
\be
ds^2 ~=~ e^{- 2 k r_c |y|} \eta_{\mu \nu} dx^{\mu} dx^{\nu} ~-~ r_c^2 dy^2
\ee
where $r_c$ is the compactification radius on a $Z_2$ orbifold, and $k \sim M_5$, the
higher dimensional Planck mass.  For the bulk KR field $B_{\mu' \nu'}$ in this scenario, one can use the standard 
decomposition technique used, for example, in \cite{gold}:
\be
B_{\mu' \nu'} (x, y) ~=~ \sum_{n} \frac{B_{\mu \nu}^{(n)} (x)}{\sqrt{r_c}}~ \xi^{(n)} (y)
\ee
which, on the visible brane, is given by
\be
B_{\mu \nu} (x) ~=~ \sum_{n} \frac{B_{\mu \nu}^{(n)} (x)}{\sqrt{r_c}}~ \xi^{(n)} (\pi)
\ee
The couplings are controlled by appropriate warp factors arising from $\xi$ 
\cite{gold}. The 4-dimensional effective parity-violating
part of the Lagrangian density for KR-fermion coupling in this case is given by
\bea
{\cal L}^{~(pv)}_{f} &=& - \frac {2 q_1}{r_c^3} ~\bar{\psi}~ [i \gamma^c 
\sigma^{a b}~ e^{6 \pi k r_c} ~\eta^{\alpha \rho} \eta^{\beta \sigma} \eta^{\delta \kappa}
~ \epsilon_{\alpha \beta b c} \nonumber \\
&\sum_{n, n'}& \{ B_{\delta a}^{(n)} \partial_{[\rho} B_{\sigma \kappa]}^{(n')}~ 
\xi'^{(n)} (\pi) \xi^{(n')} (\pi) ~+~ \partial_{[a} B_{\rho \delta]}^{(n)} B_{\kappa \sigma}
^{(n')}~ \xi^{(n)} (\pi) \xi'^{(n')} (\pi) \} ] \psi \nonumber \\
&-& \frac{q_1 + 2 q_2}{2 r_c^3} ~\bar{\psi}~ [\gamma^c \sigma_{a b} \gamma_5~
e^{6 \pi k r_c} ~\eta^{a \rho} \eta^{b \sigma} \eta^{\delta \kappa} \nonumber \\
&\sum_{n, n'}& \{ B_{\delta c}^{(n)} \partial_{[\rho} B_{\sigma \kappa]}^{(n')}~ 
\xi'^{(n)} (\pi) \xi^{(n')} (\pi) ~+~ \partial_{[c} B_{\rho \delta]}^{(n)} B_{\kappa \sigma}
^{(n')}~ \xi^{(n)} (\pi) \xi'^{(n')} (\pi) \} ] \psi
\eea
where $\xi'^{(n)} (\pi) = d\xi^{(n)}/dy\mid_{(y = \pi)}$.

It should be mentioned in this context that a 5-dimensional scenario also admits of an
additional term of the form
\be
{\cal{L}}^{~pv}_{HB} ~=~ M_p~\epsilon^{\mu\nu\lambda\alpha\beta} H_{\mu\nu\lambda} 
B_{\alpha\beta}.  
\ee
Such a term is invariant under the Kalb-Ramond gauge transformation 
$\delta B_{\mu\nu} = \partial_{[\mu} \omega_{\nu]}$, modulo a divergence term.
However, it is not invariant under the $U(1)$ gauge transformation of the KR field,
which we have introduced to justify the Chern-Simons terms defined earlier.
Therefore, a term of this form survives only if torsion does not couple to electromagnetism,
at least through a Chern-Simons term. 

Once a term of this kind exists, one hopes to generate a parity violation in four dimension
when the fifth dimension is compactified a la Randall-Sundrum. However it is found that the presence of
such a term makes the $B_{\mu\nu}$ field selfdual or anti-selfdual and the resulting four dimensional 
action conserves parity. We shall report the details of such a scenario in a forthcoming paper.

\subsection{Construction of ~$^{*}H$~ in 6 dimensions}

The construction of $^{*}H$~ in 6 dimensions can
only be completely antisymmetric in all covariant(contravariant) indices 
\cite{bmssgsen}: ~
$^{*}H_{\mu' \nu' \lambda'} ~=~ \epsilon^{\alpha' \beta' \gamma'}_{~~~~~~\mu' \nu' 
\lambda'} H_{\alpha' \beta' \gamma'}$. As has been the cases in 4 and 5 dimensions, if one 
calculates here the parity-violating part of the EC-KR-Maxwell Lagrangian, i.e., the term ~
$^{*}H^{\mu' \nu' \lambda'} H_{\mu' \nu' \lambda'}$,~ it turns out to be zero again.
Moreover, for the KR-fermion coupling, it has been shown explicitly in \cite{bmssgsen} 
that although the augmentation of the covariant derivative with the pseudo-tensorial part 
in presence of torsion causes parity-violation in the bulk, the ensuing theory in 4 dimensions 
turns out to be parity-conserving. This can be understood from the fact that upon an 
ADD-type compactification, one can obtain the following KR coupling to the spin-1/2 
fermion of mass m:
\be
{\cal L}_f ~=~ {\cal L}_f^E ~+~ M_p^{-1}~ \sum_{\vec{n}} \bar{\psi}~ \left(i \gamma^\mu
\sigma^{\nu \lambda} ~\partial_{[\mu} B_{\nu \lambda]}^{\vec{n}} \right)~ \psi ~-~
\frac{144qm}{M_p}~\bar{\psi}~i \gamma_5 \chi ~\psi
\ee
where ~${\cal L}_f^E$~ is the 4-dimensional Dirac Lagrangian in Einstein gravity, 
~$q$~ being the charge of the pseudo-connection and $\chi$, the scalar field in 
the KK spectrum for $B_{\mu' \nu'}$. From the viewpoint of parity transformation 
in 4 dimensions, this Lagrangian is invariant, since we can always use the phase 
freedom of the fields $B_{\mu \nu}^{\vec{n}}$ and $\chi$ independently on 
the 3-brane. It has also been argued in \cite{bmssgsen} that the above feature of
getting no parity-violation in any sector in 6-dimensions holds in a RS framework as well.

\section{Summary and conclusions}

We have made a general survey of the role of spacetime torsion as a 
possible source of parity violation, evinced from its interaction with both
curvature and various spin fields. We have shown that while a completely 
antisymmetric torsion (originating from Kalb-Ramond field in a string 
inspired model) can  induce parity violation only in the spin 1/2 fermion 
sector but not in the curvature or $U(1)$ gauge sector. A  two-index antisymmetric torsion 
can however violate parity in all spin sectors.

We have also generalized these results into higher spacetime dimensions. 
These results are specially significant in studying parity violation 
in phenomenological models originating from D-branes. Postulating the 
existence of torsion (identified with the KR field) in the bulk in each case, 
we still find that parity is always restored when one considers its coupling 
to curvature. On the other hand, the fermionic sector is seen to violate 
parity in the resulting  4-dimensional theory obtained upon compactification 
of the extra dimensions  {\it a la} Randall-Sundrum and ADD. In each of the 
above cases, all the parity violating couplings are explicitly calculable.
These parity violating couplings may turn out to be phenomenologically 
significant in the context of solar neutrino problem \cite{ssgas}. 

We conclude with the observation that the curvature (or gravity) sector and electromagnetic sector are
always seen to be shielded from parity-violating effects whenever the torsion
tensor is fully antisymmetric in all three indices. This, in turn, is traced to
the fact that such a tensor can always be expressed in terms of its dual axion field.
Thus  parity conservation in gravity, space-time torsion notwithstanding,
has a rather striking relationship with duality.
We have thus exhaustively described the possibilities of generating parity violating interactions 
through spacetime torsion with special emphasis on a String inspired models. Various phenomenological implications of the results 
presented in this work may now be investigated for $(3+1)$ dimensional as well as higher dimensional theories.
It may be noted from equ.(4) that $q$ measures the relative strength between the parity conserving and parity 
violating part in the Cartan extension of the affine connection. Thus to determine $q$ one must look into phenomena
originating from both the parity violating and parity conserving part. Calculating the helicity flip amplitude from left handed
to right handed neutrino and the resulting change of flux of the incoming left handed solar neutrino\cite{ssgas} 
we can compare this against the 
experimental data to estimate the parity violating component. The parity conserving part does not contribute in this process.
Data from CMB anisotropy can also be used to determine the parity violating part\cite{dmpmssg}. Moreover the experimental value
of the optical rotation of the plane of polarization of the  
distant galactic polarized radiations, over and above the usual Faraday rotation ,may be used to determine both
the parity conserving as well as parity violating components\cite{dmssg}.  
For the higher dimensional theories like ADD scenario equ.(22) indicates that only the massive Kaluza-Klein tower of the
KR field contribute in KR-fermion interaction term whereas in RS scenario equ.(26) implies that both the massless as well
as the massive modes of the KR field interact with the fermions. As the massless mode in RS scenario is shown to be suppressed 
by the large warp factor on the visible brane \cite{bmsenssg}, the massive KR modes  in these higher dimensional theories
are expected to play crucial roles in the forthcoming Tev scale experiments. 
\vskip .2in \noindent
{\Large{\bf Acknowledgments}}
\vskip .1in \noindent
The work of BM and SSG is partially supported by the Board of Research in Nuclear 
Sciences, Government of India, under grant Nos. 2000/37/10/BRNS, 98/37/16/BRNS 
cell/676. S. Sur acknowledges support from the Council of Scientific and
Industrial Research, Government of India. The work of S.S is financed by Funda\c c\~ao para a Ci\^encia e a Tecnologia, Portugal, through CAAUL.



\end{document}